\documentstyle[11pt,aaspp]{article}
\def\eg{{\it e.\thinspace g.}}
\def\etal{{\it et al.}}
\def\gcc{\hbox{\rm\hskip.35em  g cm}$^{-3}$}
\def\ergsec{\hbox{\rm\hskip.35em erg s}^{-1}}
\begin{document}
\title{ Frictional Heating and Neutron Star Thermal Evolution }
\author{Kenneth A.\ Van Riper,
Bennett Link$^1$\altaffiltext{1}{Also Department of Physics, Montana
State University, Bozeman, Montana 59717.}, and Richard I. Epstein}
\affil{Los Alamos National Laboratory,
       Los Alamos, NM 87545, USA}

\begin{abstract}

Differential rotation between the neutron star crust and a more
rapidly rotating interior superfluid leads to frictional heating that
affects the star's long-term thermal evolution and resulting surface
emission.  The frictional heating rate is determined by the mobility
of the vortex lines that thread the rotating superfluid and pin
to the inner crust lattice. If vortex pinning is relatively strong, a
large velocity difference develops between the inner crust superfluid
and the crust, leading to a high rate of heat generation by friction.
Here we present the results of thermal evolution simulations based on
two models of the vortex pinning forces that bracket a range of
plausible pinning strengths. We include the effects of superfluidity,
magnetic fields, and temperature gradients. As representative standard
and accelerated neutrino emission processes taking place in the core,
we consider the modified Urca process in normal baryonic matter, and
the much faster quark Urca process. Comparison of our
results with neutron star surface temperature data, including the
recent temperature measurement of the Geminga pulsar, shows that stars
with soft equations of state and modest frictional heating are in
closest agreement with the data; stars with stronger frictional
heating have temperatures inconsistent with the upper limit of PSR
1929+10. Stiffer stars undergoing standard cooling generally have
temperatures lying above the Vela detection, a situation worsened by
the inclusion of frictional heating. Stars undergoing accelerated
cooling without frictional heating have temperatures that fall far
below most temperature measurements; the Vela and Geminga detections
being the most compelling examples. Only in stiff stars, which have
thick crusts, can the inclusion of strong frictional heating raise the
temperature at late stages in the evolution to a level consistent with
the data. However, such a large amount of heating leads to a
temperature at $\sim 1000$ yr in excess of the Crab upper limit.
Suppression of accelerated neutrino emission processes, perhaps by
superfluid pairing in the core, may yield acceptable cooling models.

    \end{abstract}

\section  {INTRODUCTION}

The study of neutron star thermal evolution affords a means of probing
the properties of dense matter. A newborn neutron star is very hot,
with a temperature well above an MeV. The star cools initially through
neutrino emission, and later (after $\sim 10^4$ yr) cools primarily
through the emission of photons from the surface. The neutrino
emission process occurring in the core depends on the state of matter
there. In a core composed of normal baryons, neutrino emission can
occur through the relatively slow modified Urca process (``standard''
cooling), or, if the core proton fraction is high enough, through the
much faster direct Urca process (``accelerated'' cooling).  Possible
states of the core include condensed pions or kaons, bulk quark
matter, or hyperon-rich matter; neutrino emission processes in each of
these possible states would have accelerated rates comparable to the
direct Urca process.  Superfluid pairing of the nucleons is expected
to occur below temperatures of $\sim$ 1 MeV, and could have a dramatic
effect on neutron star thermal evolution by reducing the matter's
specific heat, and, depending on the core emission processes that
actually occur, the neutrino emissivity.  Furthermore, the presence of a
neutron superfluid in the inner crust would lead to frictional
heating, possibly raising the star's surface temperature significantly.

Generally, stellar evolution models simulating standard cooling ({\it
i.e.}, dominated by the modified Urca process) based on moderately
stiff to very stiff equations of state predict surface temperatures
greater than observed; whereas models simulating the accelerated
neutrino emission from an exotic or proton-rich core predict late-time
surface temperatures that are far too low (see {\it e.g.,} Friman \&
Maxwell 1979 and Van Riper 1991). Internal heating, associated with
friction between the normal matter of the inner crust and a more
rapidly rotating neutron superfluid, might allow agreement with data
for models involving accelerated cooling, while exacerbating the
problem in models of standard cooling.

Frictional heating has been examined by several authors.  Alpar \etal\
(1987) assumed an approximate balance between frictional heating and
photon emission from the stellar surface, and compared the predicted
surface temperature with the EXOSAT upper limit on the temperature for
PSR 1929+10; they concluded that only a small degree of
frictional heating is allowed.  Shibazaki
\&\ Lamb (1989) explored the effects of internal heating on the
thermal evolution of isothermal stars older than $10^3$ yr, including
possible magnetic field evolution.  Taking the internal heating rate
as a free parameter, they compared their results with EINSTEIN and
EXOSAT detections and upper limits. They found standard cooling in a
stiff star to be inconsistent with the Vela detection, and consistent
with other pulsars for various values of the heating rate. Accelerated
cooling was found to predict temperatures falling below the surface
temperatures of several pulsars, even for unrealistically high heating
rates.  Van Riper (1991a) considered constant internal energy sources,
as would occur for steady state frictional heating, in nonisothermal
neutron stars.  Van Riper, Epstein, \& Miller (1991) showed that the
heat generated during a pulsar glitch could raise the stellar surface
temperature over time scales of hours to months. Umeda, Shibazaki,
Nomoto, \& Tsuruta
\etal\ (1993, hereafter USNT) used the parametrized heating rate of
Shibazaki \& Lamb (1989) in nonisothermal simulations, and compared
their results with ROSAT data that has recently become available. They
found that standard cooling in stiff to medium stars gives
temperatures that exceed those of PSR 0656+14 and Vela.  Moreover,
strong heating in a stiff star was ruled out by PSR 1055-52, and
possibly by the PSR 1929+10 upper limit.  The study of USNT is the
most complete investigation to date, and we compare our results to
theirs below.

The purpose of our study is to consider the consequences of inner
crust frictional heating on neutron star thermal evolution by using a
microscopic description of the superfluid coupling to the crust,
rather than a parametrized form. In our analysis we use the most
recent ROSAT data of cooling neutron stars, including the Geminga
pulsar detection. We investigate a standard cooling model in which the
modified Urca process on nucleons is the dominant cooling process in
the core.  As an example of an accelerated emission mechanism, we
consider the quark Urca process.  Much recent work has focused on the
microscopic physics determining the coupling between the inner crust
superfluid and the rest of the star (Alpar 1977; Pines
\etal\ 1980; Alpar \etal\ 1984; Epstein \&\ Baym 1988; Bildsten \&\
Epstein 1989; Link
\&\ Epstein 1991, hereafter LE; Link, Epstein, \&\ Baym 1993, hereafter
LEB; Chau \&\ Cheng 1993a, 1993b).  In this study, we examine internal
heating using the microscopic description for the neutron superfluid
dynamics developed in LE and LEB. In this theory, the frictional
heating rate depends on the stellar temperature and the strength of
the interactions that pin superfluid vortex lines to the stellar
crust.

Our simulations show that accelerated cooling with frictional heating
is inconsistent with the data for both young and old pulsars; internal
heating occurring at a rate consistent with the temperatures of old
pulsars produces overheating in the early states of
evolution. Standard cooling with modest frictional heating of a very
soft star ({\it i.e.,} with a maximum mass of near 1.4 $M_\odot$) is
marginally consistent with all current temperature data. Stronger
frictional heating produces a temperature at the age of PSR 1929+10 in
excess of the upper limit for that pulsar.

This paper is organized as follows. In \S 2 we discuss the input
physics of our models; we derive the frictional heating rate, and
describe the two vortex pinning models used. In \S 3 we describe our
numerical simulations.  We compare our results with those of other
authors and with data from the EINSTEIN, EXOSAT and ROSAT satellites
in \S 4, and in \S 5 we present our conclusions. Appendix A gives the
vortex creep equations from the theory of LEB. Appendix B contains a
comparison of our time-dependent source calculations with simulations
of cooling with time-independent sources.

\section {INPUT PHYSICS}

\subsection {Cooling Processes}

A neutron star cools by neutrino emission from the interior and photon
emission from the surface.  During the first $10^4$ to $10^6$ years,
the star cools mainly through neutrino emission.  Later, surface
photon radiation dominates. The neutrino cooling rate of a neutron
star is determined by the unknown state of matter in the core.
Depending on the mass of the star and the properties of matter at
several times nuclear density, nuclear matter may extend to the center
of the star, or there may exist a distinct inner core consisting of a
pion or kaon condensate, hyperon-rich matter, bulk quark matter, or
some other exotic state. In standard cooling models, the neutrino
emission is mainly through the modified Urca process taking place in
the core.
Later, neutrino bremsstrahlung in the inner crust may be
important, though Pethick \& Thorsson (1994) argue that band structure
effects suppress this process.
Neutrino bremsstrahlung processes in the core also contribute to the
emission.  In contrast with standard models, in
accelerated cooling models neutrino emissivities are orders of
magnitude larger. A number of plausible accelerated cooling mechanisms
have been discussed, including, the direct Urca process on nucleons in
a proton-rich core, free quarks, pions, or kaons. We summarize these
neutrino processes in Table 1.

\subsection {Superfluid Effects}

Calculations of critical temperatures for nucleon pairing
(see, \eg, Hoffberg \etal\ 1970; Chao, Clark, \& Yang 1972; Ainsworth,
Pines, \& Wambach 1989; Wambach, Ainsworth, \& Pines 1991)
indicate that neutron stars form superfluids in their interiors
shortly after their formation. In the inner crust, above the
drip density of $4\times 10^{11}$
\gcc, free neutrons form a $^1S_0$ superfluid which flows
through a lattice of neutron-rich nuclei.  At or below nuclear
density, $\rho_o = 2.8\times 10^{14}$ \gcc, the nuclei dissolve,
forming a $^3P_2$ neutron superfluid, plus a dilute plasma of
superconducting protons and normal electrons.  (Recent work by Lorenz,
Ravenhall, \&\ Pethick [1993] and Lattimer \&\ Swesty [1991] suggests
nuclei dissolve at a lower density, $0.6\rho_o$.) Superfluid may or
may not extend to the center of the star.

\underline {\bf Neutrino Emission and the Specific Heat:}
Superfluidity reduces the phase space volume available for
excitations. Consequently, both the specific heat of the paired
component, and any neutrino emission process that involves nucleons
(see Table 1), would be suppressed. The direct Urca process on quarks
would be unaffected, unless the quarks become superfluid. The
possibility of quark pairing is unresolved; in this study we assume
that quark superfluidity does not occur and there is no suppression of
the quark Urca process.

\underline  {\bf Frictional Heating:} Friction between the normal
matter of the star and a more rapidly rotating interior superfluid
leads to heat generation which affects the star's thermal evolution.
The strength and time-dependence of the heat generation depends on the
microscopic physics of the interaction between the superfluid
and the normal matter. We now turn to a discussion of this
interaction, and derive the heating rate.

The angular momentum of the rotating neutron superfluid resides in a
velocity field containing singular regions of vorticity or {\it
vortex lines}.  The superfluid angular momentum is determined by the
density and distribution of these vortex lines, and can change only if
they move relative to the star's rotation axis.  Were the
vortex lines immobilized, by the pinning process discussed below for
example, the superfluid angular momentum would remain fixed.  The
rotational dynamics of the superfluid, and hence, the heating
associated with the coupling between the superfluid and the crust, is
determined by the motion of vortex lines.

In early studies of neutron star thermal and rotational evolution, the
core superfluid was assumed to be weakly coupled to the crust (Baym
\etal\, 1969; Greenstein 1975; Harding, Guyer, \&\ Greenstein 1978).
However, subsequent investigations of neutron star spin fluctuations
showed that at least 70\% of the total moment of inertia of a neutron
star is strongly coupled to the crust on time scales of greater than
three days (Boynton \&\ Deeter 1979; Boynton 1981).  Theoretical
studies suggest that the entire core is strongly coupled to the crust
(Alpar, Langer, \&\ Sauls 1984). Superconducting protons become
entrained in the neutron circulation about a vortex, resulting in a
supercurrent and an associated vortex magnetization.  Electrons
scatter efficiently off the vortex magnetic field, bringing the
velocity between the core neutron superfluid and the charged component
of the star into corotation over time scales of $\sim10^2-10^4$
rotation periods (Alpar \&\ Sauls 1988).  On longer time scales, the
core, which accounts for at least 90\% of the total moment of inertia
is tightly coupled to the crust.

Unlike the core, the inner crust neutron superfluid, which makes up $<
10$\% of the total moment of inertia, is only weakly
coupled to the crust, allowing an angular velocity difference to
develop between the superfluid and the crust as the pulsar spins down.
Friction between the two components produces heat.  The heating may
occur steadily, if the superfluid and the crust slow at nearly the
same rate, or suddenly, in a glitch.

We now derive the internal heating rate in terms of the velocity
difference between the superfluid and the rest of the star.
As the star slows under an external torque $N_{\rm ext}$,
the angular momentum of the inner crust superfluid, $J_s$, changes
according to
\begin{equation}
 I_c \dot\Omega_c + \dot J_s = N_{\rm ext} \equiv -I |\dot
\Omega_{\infty}|,\eqnum{1}
\end{equation}
where $I_c$ is the moment of inertia of the solid crust and the
corotating core, $\Omega_c$ is the spin rate of the crust, $I$ is the
total moment of inertia, and an overdot denotes a time derivative. The angular
momentum of the inner crust superfluid in the Newtonian approximation
is
\begin{equation}
J_s(t) =\int    d^3r\,\rho_s(r) r_p v_s (\vec r,t), \eqnum{2}
\end{equation}
where $r_p$ is the distance from the rotation axis, $\rho_s$ is the
superfluid mass density, and $v_s$ is the superfluid velocity, related
to the angular velocity by $v_s (\vec r, t) = r_p \Omega_s (\vec r,
t)$.
The superfluid velocity changes as vortices move radially.
Over evolutionary time scales,
the superfluid and the crust remain close to rotational equilibrium
such that $\dot\Omega_{s}=\dot\Omega_c=\dot\Omega_\infty$.

The total frictional heating rate is given by the rate of change of the total
rotational energy of the star, minus the rate at which work is done by the
external torque:
\begin{equation}
H (t) = -{d\over dt}\left [{1\over 2}I_c\Omega_c^2 +
{1\over 2}\int \rho_s v_s (\vec r)^2 \, d^3r \right ] -
|N_{\rm ext}|\Omega_c
= |\dot \Omega_c| \int \rho_s(r) r_p^2  (\Omega_s (\vec r) - \Omega_c)
d^3r, \eqnum{3}
\end{equation}
where we have used eqs. (1) and (2) to obtain the final form.  The integrand,
which represents the heating rate per unit volume, is greatest
near the equator and falls off near the rotational poles where $r_p$
vanishes. As we discuss below, the lag velocity between the
superfluid
and the crust, $\Delta v = r_p (\Omega_s - \Omega_c)$, is nearly
constant on spherical shells, except near the poles. Since we are
studying the thermal evolution with a one-dimensional computer code,
we average the superfluid heating on spherical shells to obtain
\begin{equation}
H (t) =  \pi^2 |\dot\Omega_c| \int dr\,\rho_s(r) r^3
 \Delta v = |\dot\Omega_c|
\int \omega_{\rm eff} dI \equiv J_{\rm eff}|\dot\Omega_c|, \eqnum{4}
\end{equation}
where $\omega_{\rm eff} = 3\pi\Delta v /8r$ is the effective lag
frequency and $dI=(4\pi/3)\rho_s(r)r^4dr$ is the Newtonian
approximation to the moment of inertia of a shell. This expression for
the total heating rate is useful for comparison with the results of
Shibizaki \& Lamb (1989) and USNT, who write the
heating rate in this form. In our thermal
evolution calculations, which allow for temperature gradients, we
calculate the local heating rate in each zone, properly weighted by
the moment of inertia with the general relativistic correction.

The spin rate and its time-derivative are needed in the
evaluation of the heating rate. We assume a magnetic dipole spin evolution,
\begin{equation}
\dot\Omega_c (t) = -K \Omega_c^3,   \eqnum{5}
\end{equation}
where $K$ is a positive constant. Integrating eq.\  (5) from a birth spin rate
$\Omega_c(0)=\Omega_0$ to $\Omega_c (t)$ gives
\begin{equation}
\Omega_c(t)={\Omega_0\over\sqrt {1 + 2K\Omega_0^2t}}. \eqnum{6}
\end{equation}
For times such that $\Omega_c(t)\ll\Omega_0$, eq.\  (6) simplifies to
\begin{equation}
\Omega_c (t) =( 2 K t )^{-1/2} \equiv 190 h \left (
{t\over 10^3
{\rm yr}}\right )^{-1/2} {\rm s}^{-1},  \eqnum{7}
\end{equation}
where  we have chosen a fiducial value of the constant $K$ that gives a
spin down rate that is typical of the pulsars examined in this
study. Each pulsar's spin evolution is characterized by a value of $h$
(see Table 2). For
the numerical solutions presented below, we use $h=1$.  For a birth period of
$\sim$ 1 ms, eq.\  (7) is a good approximation after
$\sim$ 10 yr.

The velocity lag appearing in the expression for the heating rate (eq.
[4]) is determined by the vortex mobility, since only if vortices move
radially can the superfluid velocity change.  Interaction forces
between vortices and nuclei, however, tend to {\it pin} vortices to
the lattice.  At high densities, above $\sim10^{13}$ \gcc, the
vortex--nucleus interaction is attractive, and vortices pin strongly to
nuclei (Alpar 1977; Epstein \&\ Baym 1988).  At lower densities, the
interaction is repulsive, and vortices pin weakly to the interstices
of the lattice (Epstein \&\ Baym 1988; LE).  In either region, vortex
pinning tends to maintain a local superfluid rotational velocity that
is higher than that of the crust. Following LEB, we consider the
motion of vortices by the process of {\it vortex creep}, which
proceeds by thermal or quantum activation of the vortices over their
pinning barriers.  Other mechanisms for moving vortex lines have been
discussed.  For example, Ruderman (1976; 1991) argues that in regions
where vortex pinning is sufficiently strong, the crust may crack
before the vortices unpin, and plates of crustal material will carry
vortices toward the stellar equator.  Jones (1990) has suggested that
vortex pinning never occurs and that the superfluid remains in a state
of near corotation with the crust.  In LEB it was shown, however, that
a vortex line adjusts its configuration in such a way as to be
much more strongly pinned than Jones estimated.

In this investigation, we assume that vortex creep allows the
superfluid to slow at the same steady rate as the crust. Steady-state
vortex creep, however, cannot account for all aspects of pulsar
timing, especially the observed glitches in spin rate. Glitches are
thought to reflect sudden transfers of angular momentum from the
neutron superfluid to the crust. Some pulsars, {\it e.g.} the Vela
pulsar, lose angular momentum mainly during glitches, rather than
through steady vortex creep. In such cases, steady-state vortex creep
is never achieved and the mean frictional heating rate is smaller than
we calculate here.

{}From vortex
number conservation, the equation of motion for the superfluid is (see,
{\it e.g.,} LE)
\begin{equation}
\dot\Omega_s (\vec r,t) \simeq {-2 \Omega_s (\vec r,t)\over r_p}
v_{cr}(\vec r, t).\eqnum{8}
\end{equation}
where $v_{cr}(\vec r,t)$ is the radial vortex creep rate.
In eq. (8) we have neglected gradients
in $\Omega_s$, which are small in the inner crust.
We consider cases such
that $|\Omega_s - \Omega_c|\ll \Omega_c$, and fix $\dot\Omega_s(\vec
r, t) = \dot\Omega_c$ in eq. (8) to obtain
\begin{equation}
v_{cr}(\vec r, t) \simeq - r_p {\dot\Omega_c \over 2 \Omega_c}. \eqnum{9}
\end{equation}
The vortex creep rate has the general form
\begin{equation}
v_{cr} = v_0^{} {\rm e} ^{-A/T_{\rm eff}}, \eqnum{10}
\end{equation}
where $v_0^{}$ is a microscopic velocity, $A$ is the activation energy
for a segment of vortex line to overcome its pinning barrier, and
$T_{\rm eff}$ is the effective temperature and which differs from the
stellar value at low temperatures because of quantum effects.  The
LEB result for the vortex creep rate is summarized in Appendix A.  The
activation energy depends sensitively on the vortex-nucleus
interaction energy (which varies with the stellar density) and on the
difference between the velocities of the superfluid and the crust,
$\Delta v(\vec r) \equiv r_p(\Omega_s - \Omega_c)$. The solution of
eqs. (9) and (10) for $\Delta v(\vec r)$ depends only logarithmically
on $r_p$, so that $\Delta v(\vec r)$ is nearly constant on spherical
shells as we assumed in obtaining eq. (4).

The strength of vortex pinning varies significantly with density in the
inner crust, especially near $10^{13}$\gcc, where the
character of pinning changes from interstitial to nuclear.  An
interstitially vortex samples only the long range part of
the interaction, and the pinning energies are very weak ($\sim 1$ keV).
Above $\sim 10^{13}$\gcc, where vortices pin directly to nuclei, the
pinning energies become difficult to calculate reliably, as the vortex
samples the poorly understood short-range nuclear pinning
potential.  We consider two models for pinning which differ in the
high-density region:

\begin{itemize}
\item
 {\it EB} Pinning Model: LE and  Epstein \&\ Baym
(1988) pinning parameters are used.

\item
 {\it ACP} Pinning Model: Above $\sim 1.5\times 10^{13}$
\gcc, pinning parameters are obtained in LEB from the approach of
Alpar, Cheng \&\ Pines (1989) based on the superfluid gap calculations
of Ainsworth, Pines, \&\ Wambach (1989). Elsewhere LE and Epstein \&\
Baym (1988) pinning parameters are used.
\end{itemize}

A calculation of the pinning interaction in the nuclear pinning region
requires solving the BCS equations for pairing of neutrons in the
presence of flow, both inside and outside nuclei. This problem has yet
to be solved. In the EB model pinning energies are estimated by using
Ginzburg-Landau theory; pinning energies in this model for the
high-density region of the inner crust are typically $\sim 10$ MeV. In
the ACP model, pinning energies are estimated by simple volume
arguments; pinning energies in this model are $\sim 1$ MeV or less.

Vortex creep will produce the most heat in the regions of nuclear
pinning, since those regions have the most moment of inertia and the
highest pinning energies.  In the core, the velocity difference
between components is negligible, and there is no significant
frictional heat generation. Above $\sim 1.3\times 10^{14}$ \gcc, the
nuclear spacing and the characteristic vortex dimension become
comparable.  In this region the pinning may become ``superweak", as
suggested by Alpar \etal\ (1984), or may vanish altogether as Jones
(1991) argues.  In either case, the heating from such regions would be
small and we ignore it in this investigation.

\section {SIMULATIONS}

We consider neutron stars with gravitational masses of $1.4M_{\sun}$
constructed by solving the Oppenheimer-Volkoff equation for
hydrostatic equilibrium based on the equations of state (EOSs) of
Baym, Pethick, \&\ Sutherland (1971; BPS; with the Pandharipande
[1971] model C alternative at high density), Pandharipande \&\ Smith
(1975; PS), and Friedman \&\ Pandharipande (1981; FP). The BPS model is
representative of soft EOSs which yield stellar models with high
central densities and thin, low-mass crusts.  The PS model was chosen
as an example of an extremely stiff model with low central density and
a thick crust.  The FP EOS, a representative intermediate model,
results in a crust size within the range found using current models
for the high density EOS (\eg, Prakash, Ainsworth, \&\ Lattimer 1988).
Radii $R$ and central densities $rho_c$ for $1.4M_{\sun}$ stars constructed
from
these EOSs are given in Table 3. As a guide for comparison with
Shibazaki \& Lamb (1989) and UNST, we also show the maximum
superfluid excess angular momentum, $J = \int dI\,\omega_{\rm crit}$,
and the maximum average lag, $\bar\omega = \int dI\,\omega_{\rm
crit}/\int dI$, integrated over the pinning region for our two pinning
models.

We consider two examples for the dominant core neutrino emission. In our
standard cooling model, we use the modified Urca process, suppressed
by superfluidity. As a specific example of accelerated neutrino
emissivity, we consider the quark Urca process (Kiguchi
\&\ Sato, 1981). Electron-nucleus bremsstrahlung, the dominant
neutrino emission process in the crust, is included in both cooling
models.

The frictional heating rate is obtained at each timestep as follows.
Using the pinning parameters given as a function of density in Table
7, eq. (9), coupled with eqs. (7) and (A1)--(A15), is solved to obtain
the superfluid velocity lag as a function of the local density and
temperature. We evaluate the local heating rate
$|\dot\Omega|\omega_{\rm eff}dI$ on a grid of spherical mass shells,
where $dI$ is the moment of inertia of each shell, with the general
relativistic correction. The thermal evolution is calculated using
the penta-diagonal scheme of Van Riper (1991b), which allows for
thermal gradients.  Evolutionary calculations of this type (see also,
\eg, Nomoto \&\ Tsuruta 1987) show that isothermality is not obtained
in the interior until an age of up to $10^4$ yr, depending on the
stellar model.  Before the star becomes isothermal, the calculated
luminosity can differ significantly from that found in an isothermal
model. The neutrino emissivities, heat capacities, and thermal
conductivities, and the details of the differencing scheme, solution
algorithms, and boundary conditions are given in Van Riper (1991b).

In Figs. 1--3 we show thermal evolution of neutron stars with and
without frictional heating. The right ordinates give $L_\infty$, the
bolometric luminosity seen by a distant observer, and the left
ordinates give the corresponding effective temperatures $T_s^\infty$.
Stars with thicker crusts, as illustrated by the PS model, have
greater frictional heating and correspondingly higher surface
temperatures at late times than stars with thinner crusts (\eg, BPS).
Moreover, the EB pinning model, which leads to larger lag velocities
and greater heating than the ACP model, produces larger surface
temperatures. At late times the surface emission is controlled
entirely by the heat source, and $L_{\infty}$ decreases as
$|\dot\Omega_c|\propto t^{-3/2}$ [see eqs. (4) and (7)].  The
temperature dependence of the vortex creep rate produces order unity
deviations in the heating rate from the $t^{-3/2}$ behavior, as shown
in Fig. 4 (top panel).  For comparison with the work of other authors,
we present the scaled heating rate $H (t) t^{3/2}$ at $10^7$ yr in
Table 4 for different pinning models and equations of state. In Fig. 4
(bottom panel) we show the evolution of the inner crust temperature
$T_m$.  The fall in $T_m$ after $\sim 10^3$ yr in the quark models
produces a larger superfluid lag and a correspondingly larger heating
rate.  Comparison of our full evolutionary calculations with models
that use time-independent energy generation are given in Appendix
B. There we find that when significant frictional heating is present,
the surface temperature is largely determined by balance between the
source and the combined neutrino and photon emission, and is
essentially independent of the star's thermal history.

\subsection{Comparison with USNT}

USNT applied the heating formulae of
Shibazaki \&\  Lamb (1989) to thermal evolution models.
Their heating is $H(t)=J|\dot\Omega_c|$ where $J$ is an assumed
constant differential angular momentum. For each stellar
model, three values of $J$ are considered, corresponding to ``superweak'',
``weak'', and ``strong'' pinning, where
$J_{\hbox{\rm strong}} = 10J_{\hbox{\rm weak}}
= 100J_{\hbox{\rm superweak}}$.
Magnetic dipole breaking was assumed, giving
$H \propto t^{-{3\over2}}$.

The values of $J_{\hbox{\rm superweak}}$, $J_{\hbox{\rm strong}}$, and
the scaled heating rate $H (t) t^{3/2}$ are listed in Table 5. USNT
heating rates are plotted as a function of time in Fig. 5 for an FP
star, together with our standard cooling model results. For PS stars,
which have thicker crusts, our EB heating is $\sim 3$ times stronger
than shown in Fig. 5, and for a BPS star, which has a thinner crust,
the EB heating is $\sim 2$ times less than shown. The changes for the
ACP heating among different stellar models are relatively smaller.

The results of USNT at early times ($<500$ yr) differ from ours
primarily because of the differing treatments of the early-time
heating rate. USNT adopt a nearly constant heating rate of $< 10^{38}$
erg s$^{-1}$ for times much earlier than 300 yr. However, superfluid
condensation is expected to occur in the inner crust within a year of
the star's birth, leading to substantial frictional heating at early
times.  Our treatment of early-time thermal evolution includes this
time-dependent heating. The USNT cooling curves exhibit a small $\dot
T$ at early times because their relatively low heating rate cannot
compete with the rapid neutrino emission. The USNT results exhibit a
rapid drop in surface temperature at $\sim 30$ yr caused by the
arrival at the surface of the cooling wave from the core, rather than
by the star's response to the changing heating rate. In our models,
however, the temperature evolution has nearly zero slope at early
times, when the heating and neutrino cooling rates essentially
balance. The rapid drop in temperature does not occur until after
$\sim 100$ yrs in our quark models because the initiation of the
cooling wave is delayed by our larger early-time heating rate.  At
late times, when the thermal evolution is controlled by the same
heating rate, our results are consistent with those of USNT.

\section{COMPARISON WITH OBSERVATIONS}

Soft x-ray emission has been detected from the surfaces of at least
four neutron stars, and interesting upper limits have been obtained
for numerous other objects. We now discuss the measurements, compare
these data with our cooling simulations, and obtain constraints on the
neutron star equation of state and internal heating source. The data
are shown along with our cooling simulations in Figs. 1--3, and our
results are summarized in Table 6. The interpretation of a spectrum in
terms of an effective temperature $T_s^\infty$, or equivalently, a
luminosity $L_\infty$, quantities we show in our figures, depends on
the full spectrum of radiation emitted by the star. A first estimate
of the effective temperature could be obtained by assuming that the
star radiates as a blackbody.  However, the dependence of the
atmospheric opacity on photon energy can lead to a substantial
difference between the temperature inferred from a blackbody fit
($T_{BB}$) and the effective temperature. For example, in a
nonmagnetic atmosphere composed of light elements, the opacity drops
with photon energy in the detector bandpass. As a result, the deeper,
hotter atmospheric layers are exposed at higher energies, hardening
the emitted spectrum, and the effective temperature can be up to a
factor of three less than $T_{BB}$ (Romani 1987). Strong magnetic
fields broaden opacity features, softening the spectrum relative to
the nonmagnetic case. Differences in treatment of the
polarization-dependent transport have given different results. When
considering only the portion of the spectrum in the ROSAT bandpass,
Miller (1992; 1993) obtains effective temperatures about equal to
$T_{BB}$ for H, He, C, and N atmospheres, while for the EINSTEIN
bandpass $T_{BB}$ can be significantly less than the effective
temperature --- up to a factor of 0.7 in C and N atmospheres.  On the
other hand, Shibanov \etal\ (1993) obtain effective temperatures
intermediate between Romani's (1987) nonmagnetic values and
$T_{BB}$. In a non-magnetic Fe atmosphere, the average soft x-ray
opacity is dominated by the presence of spectral lines, the spectrum
is softened relative to a light-element atmosphere, and the effective
temperature is close to $T_{BB}$ (Romani 1987). The structure of heavy
elements in strong magnetic fields is poorly understood, making an
assessment of the full consequences for the spectrum impossible at
present, however, because magnetic fields tend to soften the spectrum
it is not impossible that such an atmosphere could have an effective
temperature exceeding $T_{BB}$. Given the above caveats, in Figs. 1--3
we show the reported effective temperatures obtained from blackbody
fits.

Except for the Crab, whose age is known, we use the ages estimated from
the pulsar spin down.  This estimate is uncertain by at least a factor
of three.  For stars older than 100,000 years, varying the age within
this uncertainty may, depending on the model assumed,
bring the observational point above or below our cooling curves.
We show an error cross in the upper right of Figs. 1--3 to represent the
uncertainties in the spindown ages and in the temperatures due to
differences in atmosphere models.

\underline {PSR 0656+14:} The flux and spectrum over a limited band have been
measured for this pulsar. From EINSTEIN observations, C\'ordova \etal\
(1989) deduce $T_s^\infty$ between 3
and $6\times 10^5$ K.  Finley, \"Ogelman,
\&\ Kizilo$\breve{\hbox{\rm g}}$lu (1992) obtain the spectrum with the
ROSAT Position Sensitive Proportional Counter and find an effective
surface black body temperature of $T_s^\infty \simeq 9.0\pm 0.4\times
10^5$ K. Radiation transfer in an atmosphere could have significant
effects on the determination of $T_s^\infty$.
Finley,
\"Ogelman, \&\ Kizilo$\breve{\hbox{\rm g}}$lu (1992) obtain $\simeq 2.4\pm
0.2\times 10^5$ K using the non-magnetic He atmospheric model of
Romani (1987).  Finley \&\ \"Ogelman (1993) find $9.2\pm
0.6\times10^5$ K using the magnetic ($B=9.4\times 10^{11}$ G) He model
of Miller (1992), while Anderson, \etal\ 1993, using the magnetic
($B=4.7\times 10^{12}$ G ) H model of Shibanov \etal\ (1993), obtain
$5.35\pm 0.35\times 10^5$ K.  Complicating the interpretation even
further is a 14\%\ pulsation in the thermal X-rays, suggesting a
distribution of temperature over the surface.  We show the blackbody
results of Finley
\"Ogelman, \& Kizilo$\breve{\hbox{\rm g}}$lu (1992) in Figs. 1--3.

\underline {Vela (PSR 0833-45):} The flux and spectrum over a limited
band have been measured for this star. The EXOSAT detections
(\"Ogelman
\&\  Zimmermann 1989),
fit assuming a blackbody source spectrum, are shown in
in Figs. 1--3.

\underline {PSR 1055-52:} The flux and spectrum over a limited band
have been measured by EXOSAT for this pulsar (Brinkmann \&\ \"Ogelman
1987).  We show the reported temperature error bar derived from a
blackbody fit.

\underline {Geminga:} Halpern \& Ruderman (1993) use a two-component
model to fit the ROSAT spectrum, and obtain a high-temperature
component of $\sim 3\times 10^6$ K, attributed to a polar cap, and a
cooler component of $\sim 5.2\pm 1\times 10^5$ K, associated with the
remainder of the surface. The temperature of the cooler component is
shown in Figs. 1--3.

\underline {Crab (PSR 0531+21):} We show an upper limit to $L_\infty$
for the Crab pulsar obtained from recent ROSAT measurements by Becker
\& Aschenbach (1994).  This upper limit (and that for PSR 1929+10)
is plotted at the same $L_\infty$,
rather than at the same $T_s^\infty$, in Figs. 1--3, reflecting the
experimental limit on the flux.

\underline {PSR 1929+10:} We show an upper limit to $L_\infty$
from this pulsar obtained by Alpar, \etal\ (1987).  A recent
measurement by Yancopoulos, Hamilton, \&\ Helfand (1994) gives a
comparable upper limit based on the net luminosity of the star.
At the time and luminosity of this limit, the cooling curves are
sensitive to the surface magnetic field strength chosen.  Our curves
assume $B=10^{12}$ G.  The curves would reach the PSR 1929+10 luminosity
more quickly, by a factor of two, were $B=10^{13}$ assumed;
$B=10^{11}$ would increase the cooling time by a factor of two.

\underline {Other upper limits:} The unlabeled short
arrows in Fig. 1 are $2\sigma$ upper limits to the temperatures of
radio pulsars from the ROSAT all sky survey (Becker, Tr\"umper, \&\
\"Ogelman, 1993).  These points were derived from observed counting
rates assuming blackbody spectra.  One ROSAT upper limit, for PSR
1916+14, is labeled in the figure because it provides interesting
constraints on our models.

Searches for neutron stars have proven unsuccessful in the following
supernova remnants: Cas--A (Murray, \etal\ 1979), Kepler (Helfand,
Chanan, \&\ Novick 1980), Tycho (Gorenstein, Seward, \&\ Tucker 1983),
SN~1006 (Pye, \etal\ 1981), RCW~86, W28, (Helfand, Chanan, \&\ Novick
1980), and RCW~103 (Becker, \etal\ 1993; earlier interpretation of
EINSTEIN data by Touhy \&\ Garmire [1980] suggested a point source
with a temperature $T_s^\infty = 2.2\times10^6$ K for RCW 103.)
Present luminosity upper limits on point sources in Tycho, SN~1006,
and RCW~103 fall below the standard cooling results; the discovery of
a neutron star in any of these remnants would be strong evidence that
accelerated neutrino cooling is taking place.

\section{DISCUSSION AND CONCLUSIONS}

Frictional heating associated with differential rotation between the
inner crust neutron superfluid and the crust could significantly
affect neutron star thermal evolution, particularly at late times. Our
simulations show that frictional heating keeps the surface temperature
above $10^5$ K for $t > 10^6$ yr, perhaps to ages as great as $10^9$
yr, suggesting a population of potentially observable stars (see Figs.
1--3).  Stars with small radii and soft equations of state, as
illustrated by our BPS model, are in closest agreement with all the
current measurements and upper limits for standard cooling with ACP
frictional heating or no heating (Figs. 1--3 nd Table 6). The stronger
EB heating predicts a temperature at the age of PSR 1929+10 in excess
of the upper limit for that pulsar.  Our ACP model for a BPS star has
an excess superfluid angular momentum of $\sim 3.4\times 10^{43}$ gm
cm$^2$ rad s$^{-1}$, corresponding to an average lag of $\sim 10$ rad
s$^{-1}$.  Our conclusion differs from that of Alpar et al. (1987),
who conclude that the excess angular momentum must be less than $\sim
7\times 10^{42}$ gm cm$^2$ rad s$^{-1}$, with a corresponding average
lag of less than $\sim 0.7$ rad s$^{-1}$.  Standard cooling of a BPS
star lies slightly above the highest estimate for Vela's surface
temperature.  However, this discrepancy is not immitigable, given the
uncertainties in Vela's age and emission spectrum.  Because soft stars
have thin crusts, frictional heating is less important than in stiff
stars, and there is little difference between the cooling curves with
EB heating, ACP heating, or no heating, except at very late times.
Although soft equations of state are consistent with the neutron star
surface emission, elsewhere we have argued that an equation of state
as soft as BPS cannot account for pulsar post-glitch behavior (Link,
Epstein,
\& Van Riper 1992). In addition, the maximum stable mass of a BPS star
is only $1.414M_\odot$, which is $\sim 2$\% lower than the mass of
PSR 1913+16. A star slightly stiffer than BPS, as preferred by Brown
(1988), while consistent with both pulsar glitch data and neutron star
mass determinations, would increase the nominal discrepancy between the
cooling curves and the Vela detection.

The Vela pulsar provides the strongest argument against standard
cooling in stiffer stars, \eg\ PS or FP, which have cooling curves
lying above the Vela detection. Accelerated cooling models without
heating fall below measured surface temperatures; the Vela and Geminga
detections are the most compelling examples.  Frictional heating
lessens this disagreement, but the strongest frictional heating (\eg,
EB in a PS star; Fig. 3) barely brings the cooling into agreement with
the Vela and Geminga detections, while lying above the Crab and PSR
1929+10 upper limits. It appears that no combination of frictional
heating and accelerated cooling is consistent with current data.
Lattimer {\it et al.} (1993) and Page (1994), however, have emphasized
that accelerated cooling processes are suppressed by
superfluidity. Accounting for this suppression would give cooling
curves that fall between those of our quark models and our standard
models, and models of this sort may be consistent with current X-ray
data.

\bigskip
\centerline {\bf ACKNOWLEDGEMENTS}

It is a pleasure to thank W. Becker, F. K. Lamb, G. Pavlov, M. C. Miller, and
R.
Romani for valuable discussions. This work was performed under the
auspices of the U.S. Department of Energy.

\clearpage
\appendix
\centerline{\bf APPENDIX A. VORTEX CREEP RATE}

The vortex mobility is determined by the strength with which vortices
pin to the lattice, and by the local velocity difference $\Delta
v=r_p(\Omega_s - \Omega_c)$ between the superfluid and the crust. The
pinning parameters used in this study are listed in Table 7 for
selected densities: the pinning energy $U_0$, the vortex coherence
length $\xi$, the Wigner-Seitz lattice spacing $l_{\rm min}$, the
normalized vortex stiffness $\tau/\Lambda$, and the critical
differential velocity $v_B$ above which vortex lines cannot pin.
LEB obtain the average creep velocity for vortex lines in two limits -
the flexible limit ($\tau\ll 1$), in which vortex motion occurs by
unpinning at one site at a time, and the stiff limit $(\tau\gg
1)$, in which a vortex segment unpins $j$ bonds simultaneously.
In these limits the
creep rates are given by

\begin{equation}
v_1\simeq {2\omega_1\over\pi}\, l_{\rm min}\, {\rm e} ^{-A_1/T_{{\rm eff},1}},
                 \qquad {\rm flexible}, \eqnum{A1}
\end{equation}
\begin{equation}
v_j  \simeq {5\omega_j\over\pi}\, l_{\rm min}\, \left [{2\pi T_{{\rm eff},j}
         \over A_j}\right ]^{1/2}{\rm e}^{-A_j/T_{\rm eff,j}}
                  \qquad {\rm stiff}. \eqnum{A2}
\end{equation}
The symbols in eqs. (A1) and (A2) are defined as follows:

\begin{equation}
A_1 = U_0\Delta^{3/2}
\eqnum{A3}
\end{equation}
\begin{equation}
T_{\rm eff,j}\equiv {\hbar\omega_j\over 2}{\rm coth}{\hbar\omega_j\over
                                                   2T}
\qquad(j\ge 1)
\eqnum{A4}
\end{equation}
\begin{equation}
\omega _1\simeq {\pi\kappa\Lambda\over 4 l_{\rm min}^2}
              \qquad ({\rm flexible}\ \tau\ll 1)
\eqnum{A5}
\end{equation}
\begin{equation}
\omega_j\simeq  {\Delta^{1/2}\over 2\pi}{\kappa\Lambda\over\tau l_{\rm min}^2}
\left[1 +{\tau \pi^2 \over 2 (j + 1)^2\Delta^{1/2}} \right]
                 \qquad ({\rm stiff}\quad \tau\gg 1)\eqnum{A6}
\end{equation}
\begin{equation}
\Lambda =  0.116 - \ln \left [{\pi\xi\over (j + 1)l_{\rm min}}\right ],
\eqnum{A7}
\end{equation}
where $\Delta\equiv 1 - \Delta v/v_B$ and $\kappa=0.0019$
cm$^2$ s$^{-1}$ is the quantum of vorticity. The quantities $j$ and $A_j$
are given by
\begin{equation}
j=3.242 \tau^{1/2} \Delta^{-1/4}\eqnum{A8}
\end{equation}
\begin{equation}
A_j =  1.577 U_0 \Delta^{3/2}j,\eqnum{A9}
\end{equation}
for $\Delta < 1/4$, while for $\Delta\ge 1/4$ they are
\begin{equation}
j={2\sqrt {2\tau}\over\Delta ^{1/4}}\left [{\rm sech}^{-1}({1
\over \sqrt 3})-
{\rm sech}^{-1}\chi +{\Delta \psi\over\sqrt 3{\cal F}}\right ],
\eqnum{A10}
\end{equation}
\begin{equation}
A_j=U_0 \sqrt\tau\Delta ^{3/4}\Biggl [{18{\sqrt 2} \over 5} I \Delta
^{1/2}-{2\Delta ^{3/2}\psi ^3 \over 6^{3/2}{\cal F}}+
{3\psi\over\sqrt6 {\cal F}}\left ( {2\over
3}-{\cal F}+{2\Delta ^{3/2}\over 3}\right )\Biggr ]. \eqnum{A11}
\end{equation}
Here ${\cal F}\equiv\Delta v/v_B=1-\Delta$, and,
\begin{equation}
I\equiv 1-{5\over 2} (1-\chi^2)^{3/2} +{3\over 2}(1-\chi^2)^{5/2},
\eqnum{A12}
\end{equation}
\begin{equation}
\psi \equiv \left
(2+{3\over\Delta ^{1/2}}-{1\over\Delta ^{3/2}}\right )^{1/2},\eqnum{A13}
\end{equation}
\begin{equation}
\chi \equiv \left [{1\over 3}\left (1+{1\over\Delta ^{1/2}} \right
)\right ]^{1/2}. \eqnum{A14}
\end{equation}

The quantities $\tau$ and $j$ are functions of one another. In our
calculations of the vortex creep steady state, these quantities
are determined iteratively - a trial value of $\tau$ is chosen, $j$ is
calculated, a new value of $\tau$ is obtained, $j$ is again
calculated, etc., until
convergence.

Eqs. (A1) and (A2) are for the limiting cases of $\tau\ll 1$ and
$\tau\gg 1$. To account for intermediate stiffness, we interpolate
between the limiting cases by using the
expression

\begin{equation}
v_{\rm cr} =
     {1\over j}\left [(1 - j) v_j + v_1\right ].
\eqnum{A15}
\end{equation}

\clearpage
\appendix
\centerline{\bf APPENDIX B. COMPARISON WITH STEADY STATE SOURCE CALCULATIONS}

Here we compare the results of our evolutionary calculations with
those of models that assume time-independent energy generation in the
crust (\eg, Van Riper 1991a). We have calculated steady-state models
for an inner crust source in an FP star. In Fig. 6 we plot the surface
luminosity as a function of steady-state source strength. In Fig. 7 we
compare the thermal evolution from our frictional heating model with
the equilibrium temperature of a steady-state heating model.
Each filled circle was obtained as follows: in our full evolutionary
calculations we extract the heating rate at a selected time. We then
perform an evolutionary calculation with this constant heating rate,
until the temperature reaches a steady state. The circle represents
the resulting surface luminosity. Where the circles fall on the
cooling curves, heat generation balances the combined surface and neutrino
emissions, and the thermal history of the star is unimportant.

Cooling curves calculated assuming an isothermal interior are shown as
{\it dashed lines} in Figs. 6 and 7.  These curves coincide with the
evolutionary results for $H \le 10^{36} \ergsec$ for both standard and
accelerated cooling, indicating that the interior temperature profile
is effectively isothermal.  The cooling curve for the standard cooling
case is above the steady state points for $10^{35} \ge H \ge
10^{32}\ergsec$ ($10^4$ to $10^6$ yr).  The additional radiation is
residual thermal energy. For higher $H$, the hot crust insulates the
interior from the surface and $L$ is determined only by the heating
rate in the crust (the initial heat content in the crust is
negligible). The difference between the steady state and evolutionary
models is greatest in our BPS model, which has relatively small heating,
and is negligible for the highly heated PS star.
The surface temperature of the accelerated cooling models is mainly
due to the source once isothermality obtains.  The heat content of the
core, which has been reduced by enhanced neutrino emission,
contributes only a negligible fraction to the surface luminosity.
The quark model cooling curve shows a sudden decline between 100 and 1000
years, similar to the drop caused by a cooling wave reaching the surface in
a star undergoing accelerated cooling with no or weak crust heating.  In
our strong heating case, however, the shape of the cooling curve is
controlled by the time dependence of the source rather than by the dynamics
of conduction in the crust.

\clearpage

$$\vbox{\halign{#&\quad#&\quad$#$\quad\hfil&#\cr
\multispan4\hfil TABLE 1\hfil\cr
\multispan4\hfil NEUTRINO EMISSION PROCESSES\hfil\cr
\noalign{\medskip\hrule\smallskip\hrule\medskip}
  & &\omit\hfil  Emissivity$^a$ \hfil&  \cr
\multispan2\hfil Process\hfil & ({\rm erg}\ {\rm s}^{-1}\ {\rm cm}^{-3})
          & \qquad Reference\cr
\noalign{\medskip\hrule\medskip}
Direct\hskip 5pt Urca\dotfill& $\cases {n \rightarrow p + e^- + \bar\nu_e \cr p
+
e^- \rightarrow n + \nu_e\cr}$ & \sim 10^{27}T_9^6 & Lattimer\hskip 5pt
et\hskip 5pt al.\hskip 5pt 1991 \cr
& & & \cr
Modified\hskip 5pt Urca\dotfill& $\cases { n + n^\prime \rightarrow n^\prime +
p +
e^- + \bar\nu_e \cr n^\prime + p + e^- \rightarrow n^\prime + n +
\nu_e\cr}$ &  \sim 10^{20}T^8_9 & Friman \& Maxwell\hskip 5pt 1979 \cr
& & & \cr
Core\hskip 5pt bremsstrahlung\dotfill& $\cases {
       n + n^\prime \rightarrow n + n^\prime + \nu_e + \bar\nu_e \cr
       n + p  \rightarrow n + p  + \nu_e + \bar\nu_e \cr
       e^- + p  \rightarrow e^- + p  + \nu_e + \bar\nu_e \cr  }$
&  \sim 10^{19}T^8_9 & Friman \& Maxwell\hskip 5pt 1979 \cr
& & & \cr
Quark\hskip 5pt Urca\dotfill&  $\cases {d \rightarrow u + e^- + \bar\nu_e \cr u
+ e^-
\rightarrow d + \nu_e\cr}$ & \sim 10^{26}\alpha_cT_9^6 & Iwamoto\hskip 5pt 1980
\cr
& & & \cr
Pion\hskip 5pt condensate\dotfill& $\cases {n + \pi^- \rightarrow n + e^- +
\bar\nu_e
\cr n + e^- \rightarrow n + \pi^- + \nu_e\cr}$ & \sim 10^{26}T_9^6 &
Maxwell\hskip 5pt et\hskip 5pt al.\hskip 5pt 1977\cr
& & & \cr
Kaon\hskip 5pt condensate\dotfill& $\cases {n + K^- \rightarrow n + e^- +
\bar\nu_e
\cr n + e^- \rightarrow n + K^- + \nu_e\cr}$ & \sim 10^{24}T_9^6 &
Brown\hskip 5pt et\hskip 5pt al.\hskip 5pt 1988 \cr
& & & \cr
Crust\hskip 5pt bremsstrahlung\dotfill& $e^- + (A,Z)$\hfil &  &\cr
 & \qquad$\rightarrow e^- + (A,Z) + \nu_e +
\bar\nu_e$ & \sim 10^{20}T_9^6 & Itoh\hskip 5pt et\hskip 5pt al.\hskip 5pt
1984\cr
\noalign{\medskip\hrule\medskip}
\noalign{\hbox to 1pc{\noindent $^a$$T_9$ is the temperature in units
of $10^9$ K.\hss}}
}}$$

$$\vbox{\halign{\quad\hfil#&\quad\hfil#&#\hfil\quad\cr
\multispan3\hfil TABLE 2\hfil\cr
\multispan3\hfil SPIN DOWN PARAMETER\hfil\cr
\noalign{\medskip\hrule\smallskip\hrule\medskip}
Pulsar & $h$& \cr
\noalign{\medskip\hrule\medskip}
Crab       &     0&.7    \cr
Vela       &     1&.3    \cr
Geminga    &     2&.6    \cr
1055$-$52  &     4&      \cr
1929+10    &     8&      \cr
0656+14    &     0&.9    \cr
1916+14    &     0&.3    \cr
\noalign{\medskip\hrule}
}}$$
\bigskip

$$\vbox{\halign{#\hfil&\quad#\hfil&\hfil\quad#\hfil&\quad\quad\quad#\quad\hfil&\hfil\quad#\hfil\cr
\multispan5 \hfil Table 3\hfil\cr
\multispan5 \hfil $M=1.4 M_\odot $ NEUTRON STARS \hfil\cr
\noalign{\medskip\hrule\smallskip\hrule\smallskip}
    &                      &                     \cr
    &\omit\quad\hfil $R$ \hfil  &   $\rho_c$  & $J/10^{44}$ & $\bar\omega$
\cr
EOS &\omit\quad\hfil  (km)\hfil   &\omit\hfil (g cm$^{-3}$)
&\omit\hfil (g cm$^2$ rad s$^{-1}$) & (rad s$^{-1}$) \cr
\noalign{\smallskip\hrule\smallskip}
BPS &  7.34 &  $4.89\times 10^{15}$ & $4^a/0.34^b$ & $145^a/13^b$\cr
 FP & 10.57 &  $1.27\times 10^{15}$ & $24/2.4$ & $97/10$\cr
 PS & 15.83 &  $4.69\times 10^{14}$ & $450/19$ & $78/3.3$\cr
\noalign{\medskip\hrule\medskip}
\noalign{\hbox to 1pc{\noindent $^a$EB model\hss}}
\noalign{\hbox to 1pc{\noindent $^b$ACP model\hss}}
}}$$

$$\vbox{\halign{\quad\hfil#&\quad\hfil#&\quad\hfil$#$\hfil
                 &\quad\hfil$#$\hfil&\quad\hfil$#$\hfil\quad\cr
\multispan3\hfil TABLE 4\hfil\cr
\multispan3\hfil HEATING RATES\hfil\cr
\noalign{\medskip\hrule\smallskip\hrule\medskip}
 & &     H(t)\,t^{3/2}\;(10\ {\rm Myr})  \cr
EOS & GAP & (\hbox{erg s}^{1/2})  \cr
\noalign{\medskip\hrule\medskip}
 PS &  EB & 2.64\times10^{42}  \cr
 PS & ACP & 1.27\times10^{41}  \cr
 FP &  EB & 1.50\times10^{41}  \cr
 FP & ACP & 1.52\times10^{40}  \cr
BPS &  EB & 2.55\times10^{40}  \cr
\noalign{\medskip\hrule}
}}$$

\bigskip
$$\vbox{\halign{\quad\hfil#&#\hfil\quad&\qquad\quad\hfil#&#\hfil\quad
                &\quad\hfil$#$\hfil\quad\cr
\multispan5\hfil TABLE 5\hfil\cr
\multispan5\hfil USNT HEATING RATES\hfil\cr
\noalign{\medskip\hrule\smallskip\hrule\medskip}
\multispan2\hfil EOS\hfil
&\multispan2\quad\hfil$J/10^{44}$\hfil\quad& H(t)t^{3/2}\;(10\ {\rm Myr}) \cr
\multispan2\hfil Pinning$^\dagger$\hfil
    &\multispan2\quad\hfil(g cm$^2$ rad s$^{-1}$)\hfil\quad&
(\hbox{erg s}^{1/2}
)          \cr
\noalign{\medskip\hrule\medskip}
PS&/S          &    31.&0  &  3.7\times10^{41} \cr
PS&/SW         &   0.&31   &  3.7\times10^{39} \cr
FP&/S          &    7.&3   &  6.4\times10^{40} \cr
FP&/SW         &   0.&073  &  6.4\times10^{38} \cr
BPS&/S         &    4.&8   &  2.9\times 10^{40} \cr
BPS&/SW        &   0.&048  &  2.9\times10^{38} \cr
\noalign{\medskip\hrule\medskip}
\noalign{\hbox to 1pc{\noindent $^\dagger$S = strong pinning, SW =
superweak pinning.\hss}}
}}$$

$$\vbox{\halign{\quad#\hfil&\quad#\hfil&
           \qquad\hfil#\hfil&\quad\hfil#\hfil&
           \qquad\hfil#\hfil&\quad\hfil#\hfil&
           \qquad\hfil#\hfil&\quad\hfil#\hfil \quad\cr
\multispan8\hfil TABLE 6\hfil\cr
\multispan8\hfil ALLOWED CASES\hfil\cr
\noalign{\medskip\hrule\smallskip\hrule\bigskip}
        &          & \multispan6\hfil Model \hfil\quad\cr
        &          & \multispan6\quad\hrulefill  \quad\cr
\noalign{\medskip}
Pulsar  & Heating  & BPS & BPS(accel)  &   FP   & FP(accel)   &   PS  &
PS(accel)\cr
\noalign{\medskip\hrule\medskip}
Crab    &  none &  Y &    Y         &     Y  &   Y         &     N &  Y  \cr
        &  ACP  &  Y &    Y         &     Y  &   Y         &     N &  Y  \cr
        &  EB   &  Y &    Y         &     N  &   Y         &     N &  N  \cr
\noalign{\smallskip}
Vela    &  none &  N &    N         &     N  &   N         &     N &  N  \cr
        &  ACP  &  N &    N         &     N  &   N         &     N &  N  \cr
        &  EB   &  N &    N         &     N  &   N         &     N &  Y  \cr
\noalign{\smallskip}
Geminga &  none &  Y &    N         &     Y  &   N         &     Y &  N  \cr
        &  ACP  &  Y &    N         &     N  &   N         &     N &  N  \cr
        &  EB   &  Y &    N         &     N  &   N         &     N &  Y  \cr
\noalign{\smallskip}
1055-52 &  none &  Y &    N         &     Y  &   N         &     N &  N  \cr
        &  ACP  &  Y &    N         &     Y  &   N         &     Y &  N  \cr
        &  EB   &  Y &    N         &     Y  &   N         &     N &  N  \cr
\noalign{\smallskip}
1929+10 &  none &  Y &    Y         &     Y  &   Y         &     Y &  Y  \cr
        &  ACP  &  Y &    Y         &     N  &   N         &     N &  N  \cr
        &  EB   &  N &    N         &     N  &   N         &     N &  N  \cr
\noalign{\smallskip}
0656+14 &  none &  Y &    N         &     Y  &   N         &     Y &  N  \cr
        &  ACP  &  Y &    N         &     Y  &   N         &     Y &  N  \cr
        &  EB   &  Y &    N         &     N  &   N         &     N &  N  \cr
\noalign{\smallskip}
1916+14 &  none &  Y &    Y         &     N  &   Y         &     Y &  Y  \cr
        &  ACP  &  Y &    Y         &     N  &   Y         &     N &  Y  \cr
        &  EB   &  N &    Y         &     N  &   Y         &     N &  Y  \cr
\noalign{\medskip\hrule}
}}$$

$$\vbox{\halign{\quad\hfil#\hfil&\quad\hfil#&#\hfil&\quad\hfil#&#\hfil
&\quad\hfil#\hfil&\quad\hfil#&#\hfil&\quad\hfil$#$\hfil\quad\cr
\multispan9\hfil TABLE 7\hfil\cr
\multispan9\hfil NUCLEAR PINNING PARAMETERS\hfil\cr
\noalign{\medskip\hrule\smallskip\hrule\medskip}
LOG$_{10}\> \rho$& \multispan2\quad\hfil $U_0$ \hfil &
\multispan2\quad\hfil $\xi$ \hfil & $l_{\rm min}$
& \multispan2\quad\hfil  $\tau/\Lambda$ \hfil& v_B^{}  \cr
(\gcc)& \multispan2\quad\hfil (MeV) \hfil &
\multispan2\quad\hfil (fm) \hfil & (fm)
& \multispan2\quad\hfil \hfil  & (\hbox{cm s}^{-1})  \cr
\noalign{\medskip\hrule\medskip}
\multispan9\hfil EB Pinning Model\hfil\cr
\noalign{\medskip\hrule\medskip}
11.83 &  0.&00066 &  2.&6 & 98. & 21.&     &    1.4\times10^5\cr
11.99 &  0.&0018  &  2.&3 & 93. & 20.&     &    1.6\times10^5\cr
12.18 &  0.&0036  &  2.&0 & 89. & 17.&     &    1.9\times10^5\cr
12.41 &  0.&0090  &  1.&7 & 84. & 16.&     &    2.1\times10^5\cr
12.79 &  0.&025   &  1.&5 & 79. & 13.&     &    2.7\times10^5\cr
12.98 &  0.&4     &  1.&4 & 71. &  0.&053  &    1.6\times10^7\cr
13.18 &  6.&4     &  1.&5 & 66. &  0.&0068 &    1.6\times10^8\cr
13.53 & 15.&0     &  1.&9 & 57. &  0.&009  &    3.8\times10^7\cr
13.89 &  9.&0     &  4.&6 & 39. &  0.&070  &    1.7\times10^8\cr
14.12 &  5.&4     & 18.&  & 29. &  0.&12   &    5.2\times10^7\cr
\noalign{\medskip\hrule\medskip}
\multispan9\hfil ACP Pinning Model\hfil\cr
\noalign{\medskip\hrule\medskip}
11.83 & 0.&00066  &  2.&6 & 98. &  21.&     &    1.4\times10^5\cr
11.99 & 0.&0018   &  2.&3 & 93. &  20.&     &    1.6\times10^5\cr
12.18 & 0.&0036   &  2.&0 & 89. &  17.&     &    1.9\times10^5\cr
12.41 & 0.&0090   &  1.&7 & 84. &  16.&     &    2.1\times10^5\cr
12.79 & 0.&025    &  1.&5 & 79. &  13.&     &    2.7\times10^5\cr
12.98 & 0.&4      &  1.&4 & 71. &   0.&053  &    1.6\times10^7\cr
13.18 & 6.&4      &  1.&5 & 66. &   0.&0068 &    1.6\times10^8\cr
13.53 & 1.&0      &  7.&9 & 57. &   0.&22   &    8.0\times10^6\cr
13.89 & 0.&97     & 12.&  & 39. &   1.&8    &    3.2\times10^6\cr
14.12 & 0.&22     & 30.&  & 29. & 110.&     &    2.4\times10^5\cr
\noalign{\medskip\hrule}
}}
$$

\begin{figure}

\caption{Thermal evolution of an FP star of gravitational mass $1.4M_{\sun}$
without frictional heating ({\it solid curves}) and with heating using
the EB ({\it dashed curves}) and ACP ({\it dot--dashed curves})
pinning parameters.  For each case, the evolution was calculated with
``standard'' neutrino emissivities and with ``accelerated'' cooling due to
emission from quarks.  The {\it dotted curve} is an isothermal
approximation to the standard EB model.  A surface magnetic field of
$10^{12}$ G is assumed. Data and upper limits, discussed in the text,
are shown. Uncertainties associated with the atmospheric physics and age
determinations are shown by the error cross on the right side of each
figure (see text). When comparing theoretical curves with data, two factors
must be borne in mind. 1) For all pulsars except the Crab, the ages
shown correspond to a breaking index of 3; an index of 2 gives half
the age shown, while an index of 4 give 1.5 times the age shown. 2)
The theoretical curves correspond to an $h=1$ spin-down parameter (see
eq. [7]) The heating rate of a given pulsar is proportional to its
spin-down parameter (given in Table 2).  At late times, the luminosity
equals the heating rate, and the curves with heating must be shifted by
a factor $h$ in the luminosity. At early times, the required shift can be
estimated by the difference between the EB and ACP curves, which
differ by a factor of $\sim 10$ in the heating rate.}

\caption{Same as Fig. 1, for a PS star.}

\caption{Same as Fig. 1 for a BPS star, except accelerated
cooling with heating was calculated assuming an isothermal core to
save computer time.}

\caption{Evolution of the scaled heating rate, $H(t)t^{3/2}$,
normalized to its value at at $10^7$ yr ({\it top panel}), and the
inner crust temperature ({\it bottom panel}).}

\caption{Net heating rates for an FP star from this
work ({\it solid lines}) and from USNT ({\it dashed lines}) for several
pinning models.}

\caption{Steady state surface luminosity  as a function of inner crust
source rate for FP neutron stars. The luminosity is equal to the
source rate along the {\it dotted line}.}

\caption{Cooling curves for an FP star compared with steady state results
                         ({\it filled circles}).
The logarithm of the source strengths $H$ in $\ergsec$ is shown above
the bottom
axis. Models calculated assuming isothermality are shown as {\it dashed
lines}.}

\end{figure}
\clearpage
\begin{figure}

\end{figure}
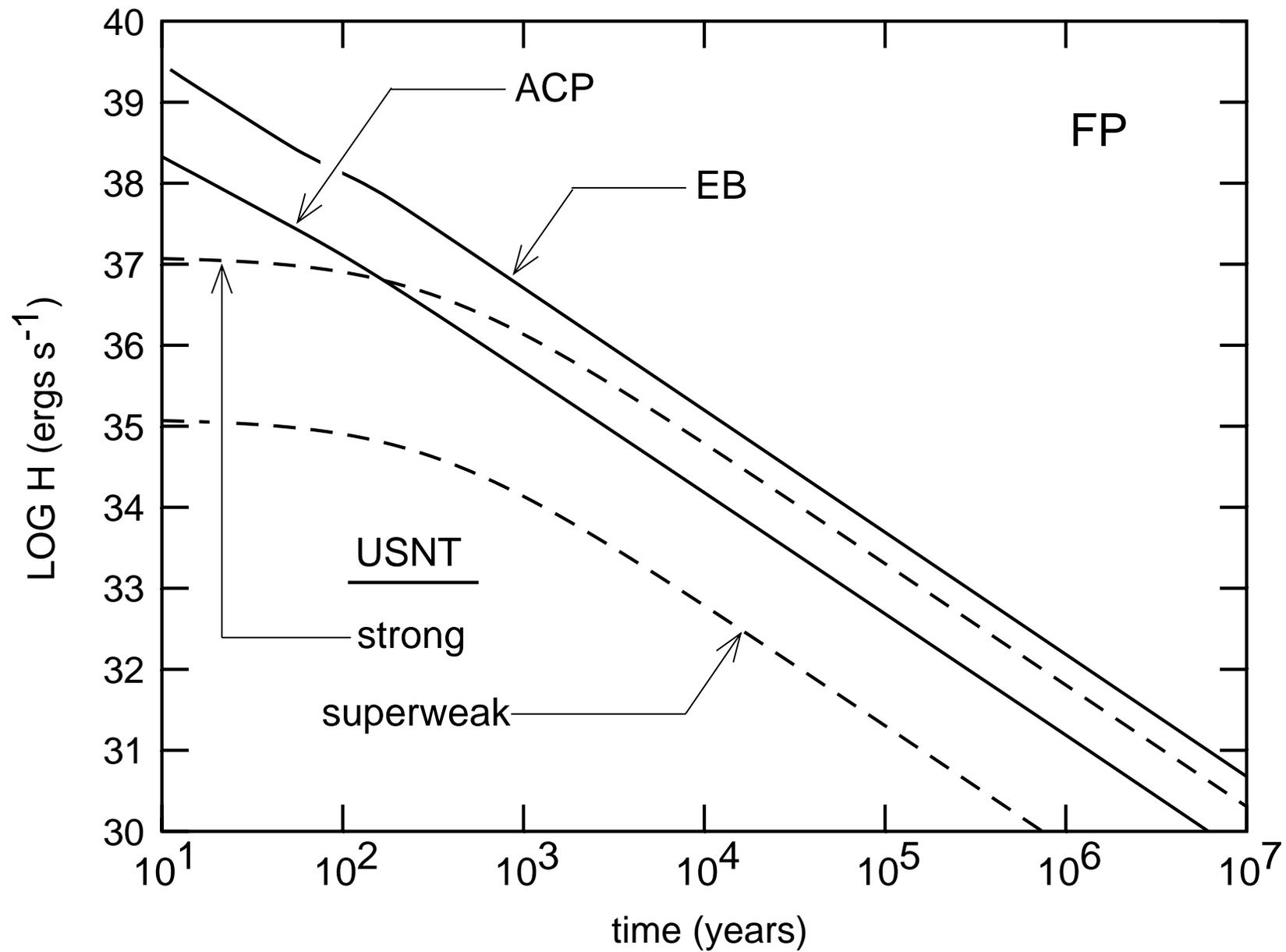
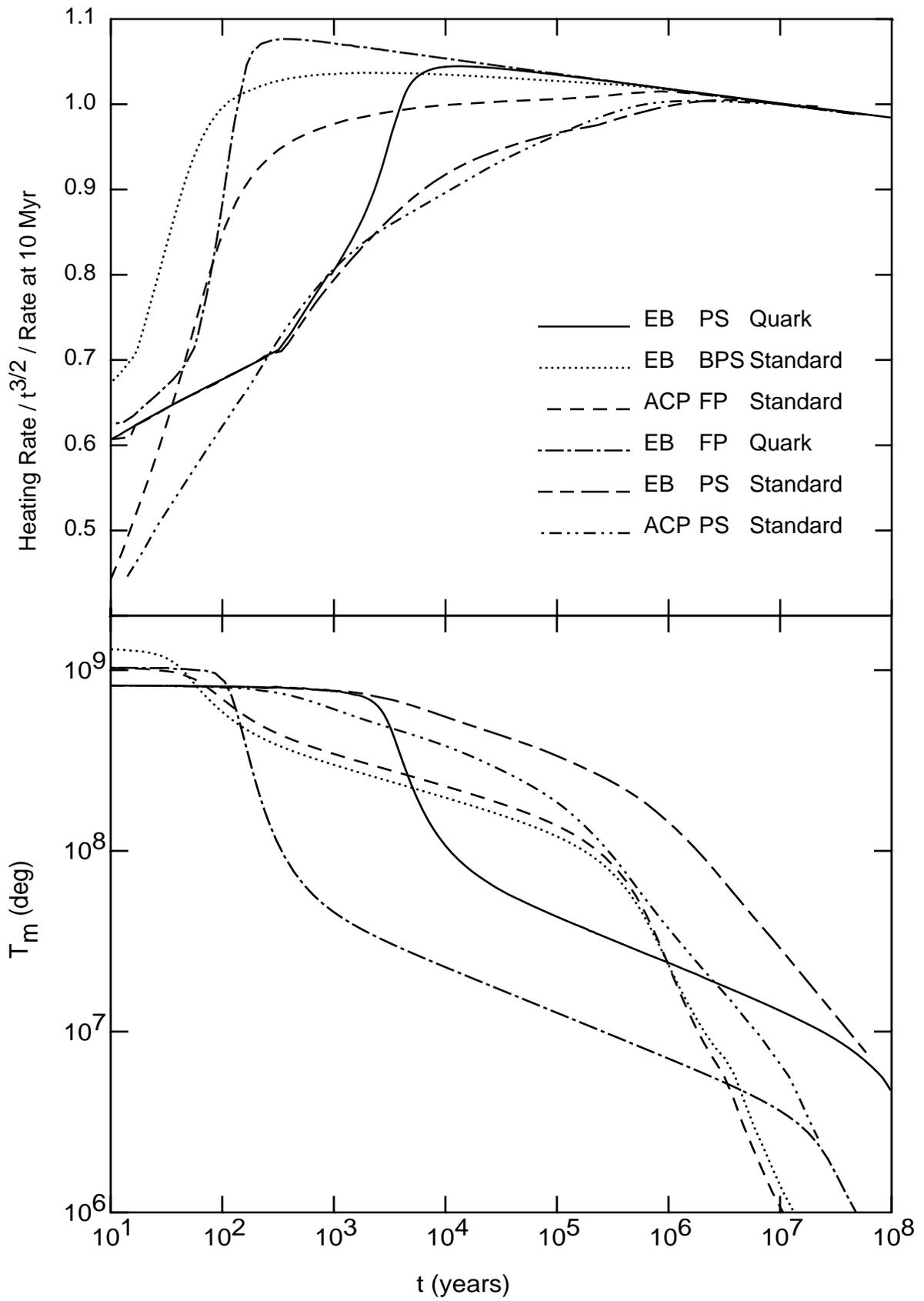
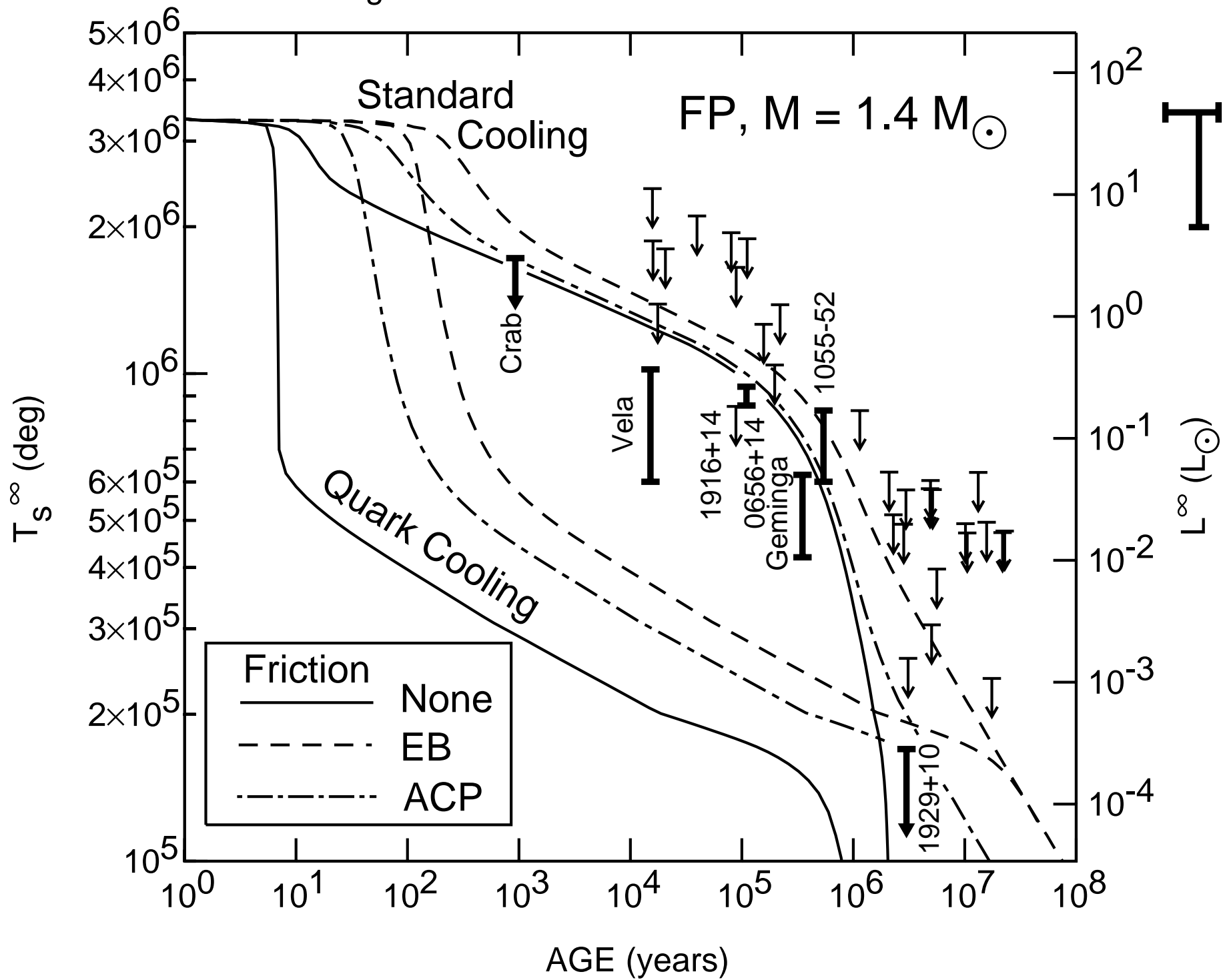
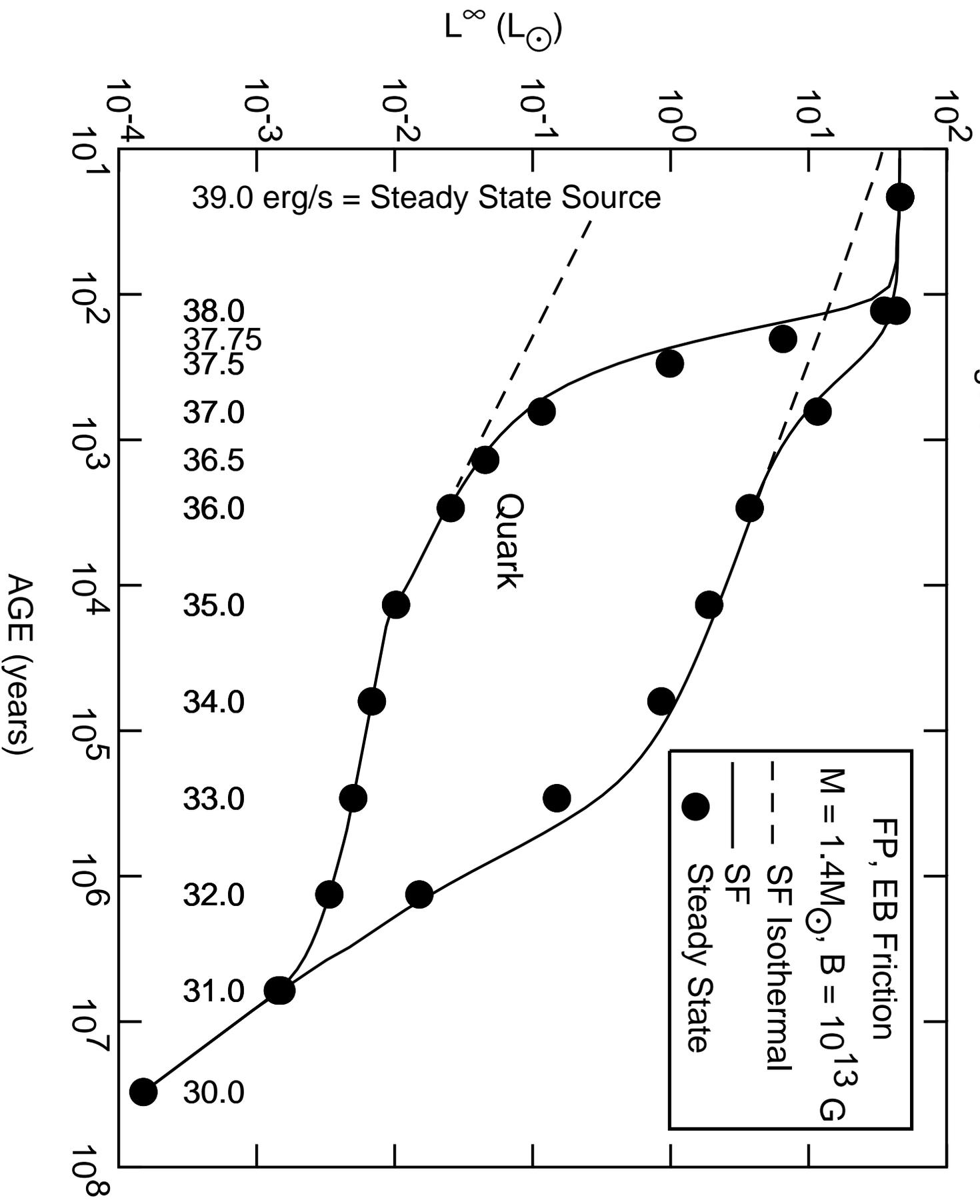
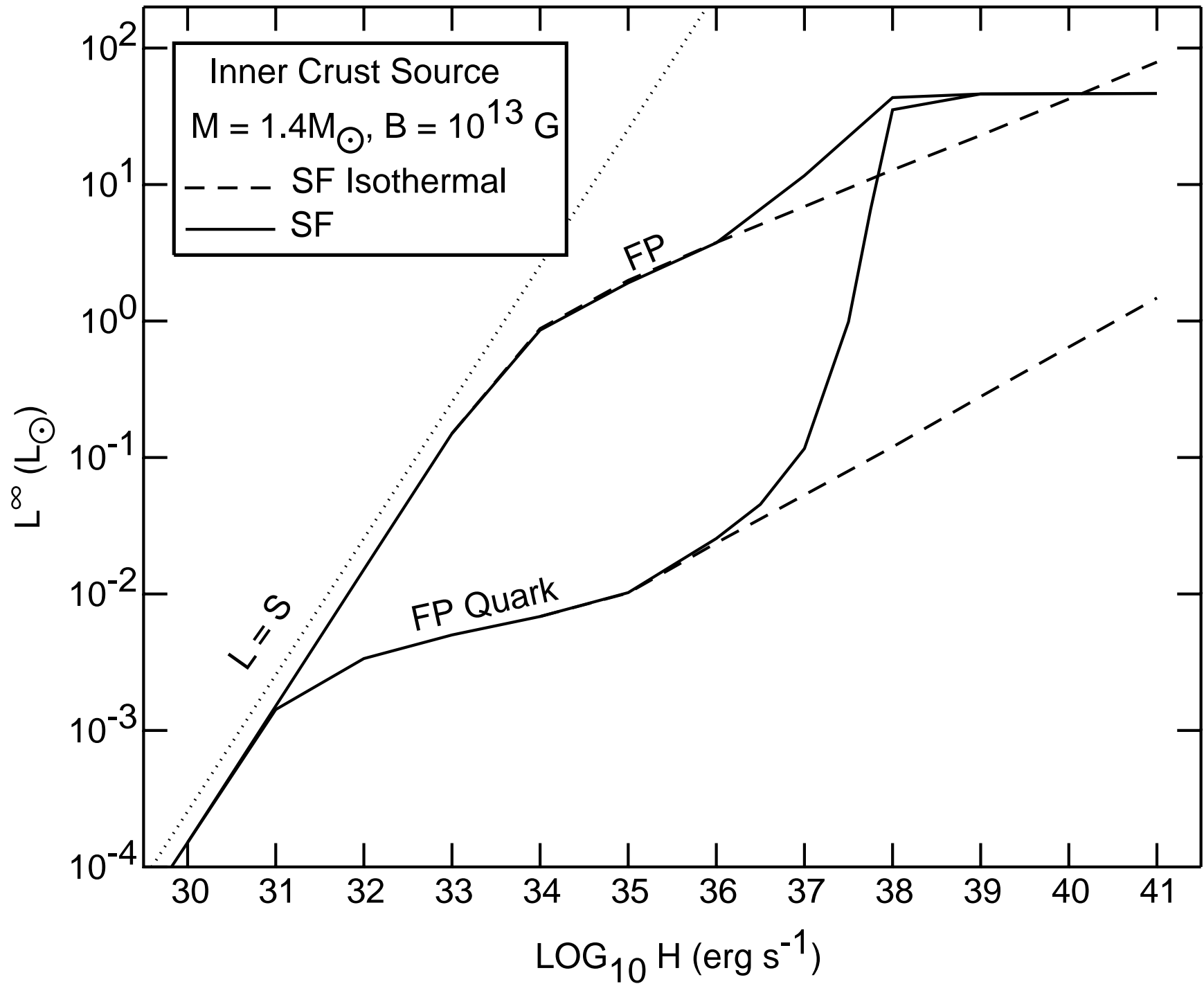
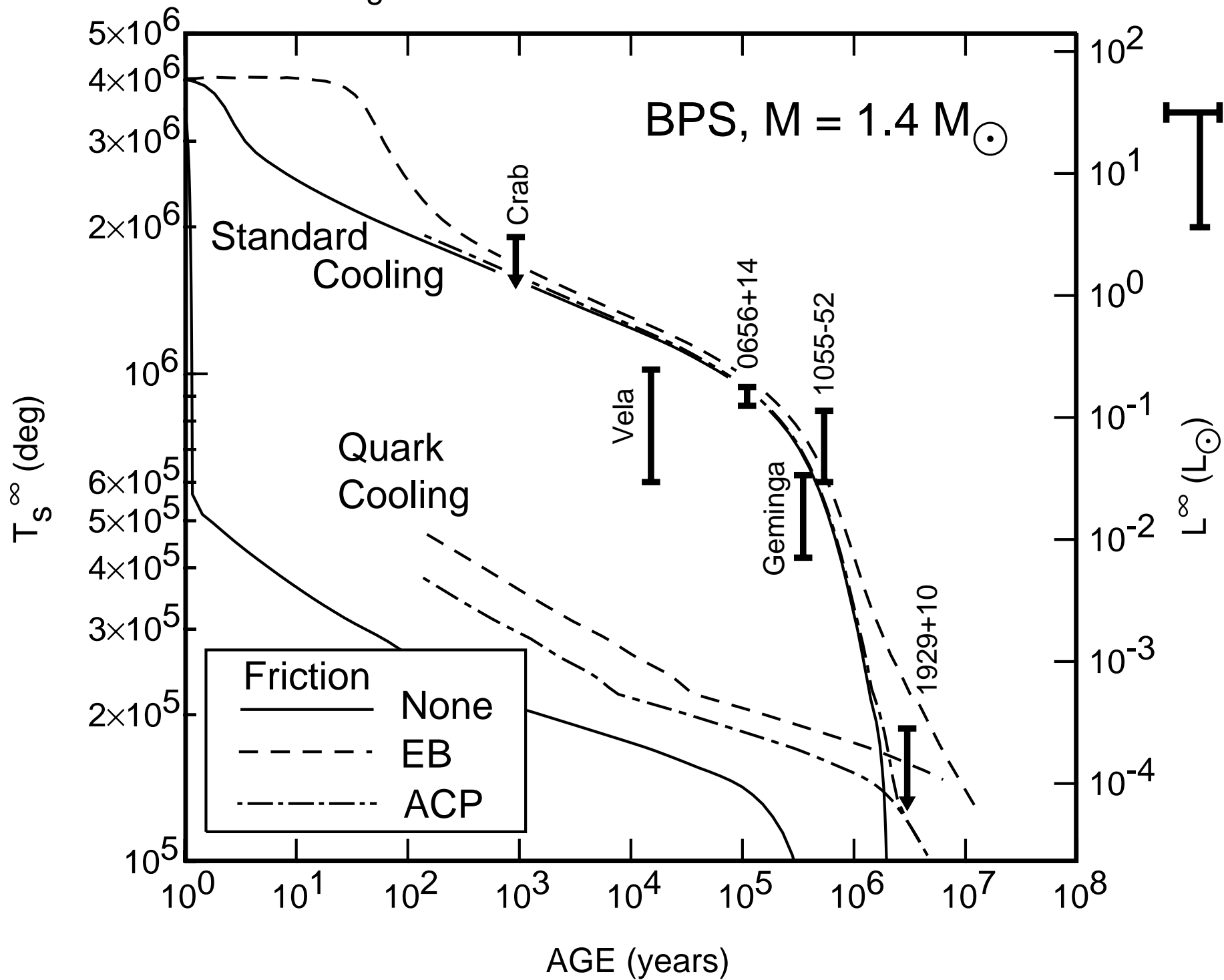
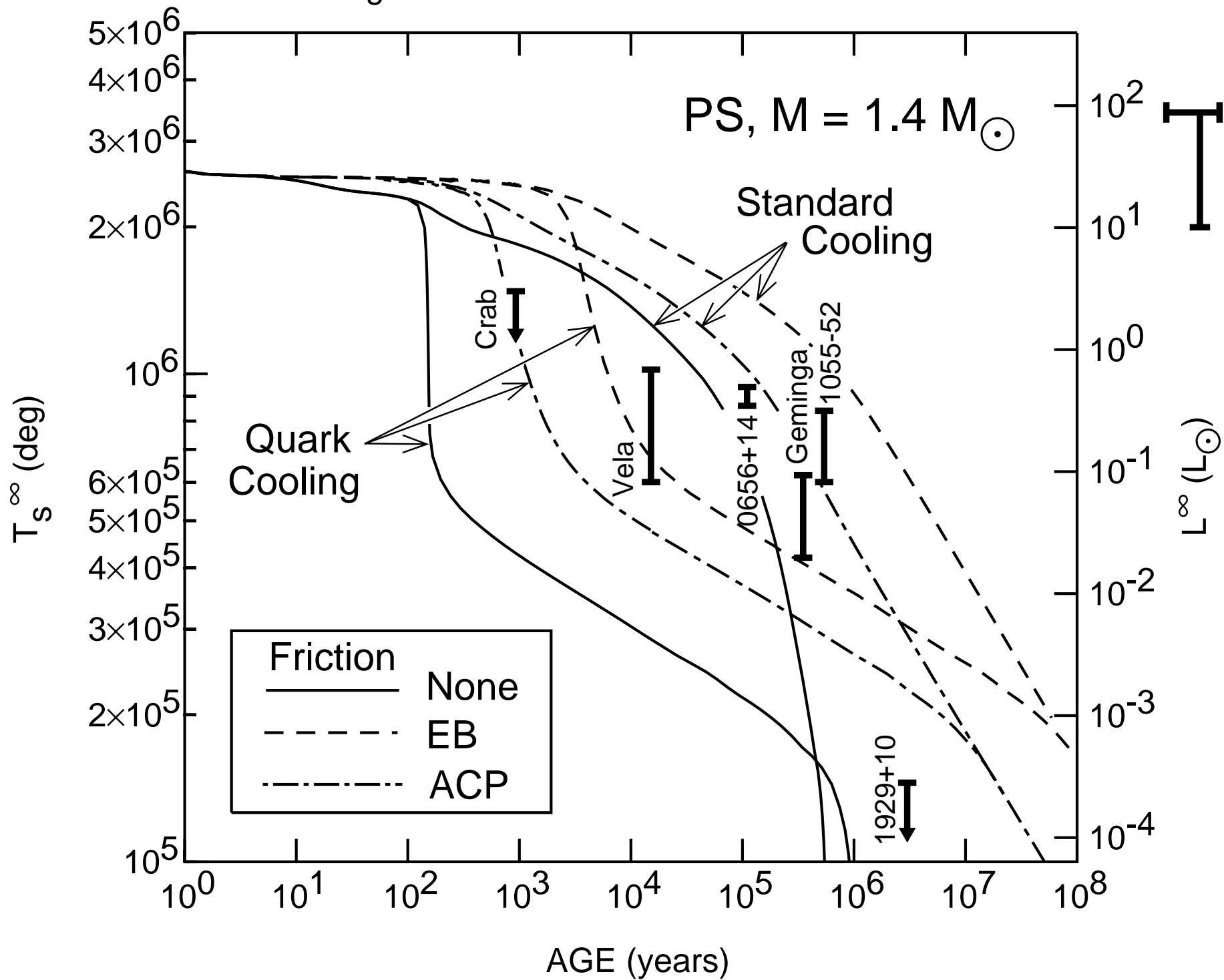
\end{document}